\documentclass[preprint,pre,longbibliography]{revtex4-1} 

\usepackage{natbib}
\usepackage{graphicx}
\usepackage{amssymb}
\usepackage{amsfonts}
\usepackage{amsmath}
\usepackage{amsbsy}
\usepackage{mathrsfs} 
\usepackage{color}
\usepackage{gensymb} 
\usepackage{hyperref} 
\usepackage{soul}%\st{...}
\usepackage{esint}
\usepackage{mymacros}
\def\XXint#1#2#3{{\setbox0=\hbox{$#1{#2#3}{\int}$}
     \vcenter{\hbox{$#2#3$}}\kern-.5\wd0}}

\begin{document} 

\title{Acoustics of a partially partitioned narrow slit connected to a half-plane: case study for exponential quasi-bound states in the continuum and their resonant excitation} 

\author{Ory Schnitzer}
\affiliation{Department of Mathematics, Imperial College London, 180 Queen's Gate, London SW7 2AZ, UK}

\author{Richard Porter}
\affiliation{School of Mathematics, University of Bristol, Woodland Road, Bristol BS8 1UG, UK}

\begin{abstract}
Localised wave oscillations in an open system that do not decay or grow in time, despite their frequency lying within a continuous spectrum of radiation modes carrying energy to or from infinity, are known as bound states in the continuum (BIC). Small perturbations from the typically delicate conditions for BIC almost always result in the waves weakly coupling with the radiation modes, leading to leaky states called quasi-BIC that have a large quality factor. We study the asymptotic nature of this weak coupling in the case of acoustic waves interacting with a rigid substrate featuring a partially partitioned slit --- a setup that supports quasi-BIC that exponentially approach BIC as the slit is made increasingly narrow. In that limit, we use the method of matched asymptotic expansions in conjunction with reciprocal relations to study those quasi-BIC and their resonant excitation. In particular, we derive a leading approximation for the exponentially small imaginary part of each wavenumber eigenvalue (inversely proportional to quality factor), which is beyond all orders of the expansion for the wavenumber eigenvalue itself. Furthermore, we derive a leading approximation for the exponentially large amplitudes of the states in the case where they are resonantly excited by a plane wave at oblique incidence. These resonances occur in exponentially narrow wavenumber intervals and are physically manifested in cylindrical-dipolar waves emanating from the slit aperture and exponentially large field enhancements inside the slit. The asymptotic approximations are validated against numerical calculations. 
\end{abstract}

\maketitle

\section{Introduction}
In an open system, unforced wave oscillations normally decay with time even in the absence of dissipation. This happens because of radiation damping, whereby local oscillations couple to waves propagating to infinity. There are rare counter examples, however, of localised time-harmonic waves that co-exist, yet do not couple with outgoing propagating waves. These are called bound states in the continuum (BIC), embedded trapped modes, edge states or dark states; they were first described in quantum mechanics by von Neumann and Wigner \cite{VN:29,Stillinger:75} and since established as a general wave phenomenon with potential applications across wave physics, especially in acoustics, water waves and photonics \cite{Evans:94,Linton:07,Hsu:16,Koshelev:19}. The terminology alludes to the fact that BIC occur at single isolated frequency eigenvalues embedded in the continuous frequency intervals in which waves are allowed to propagate to and from infinity; they are distinguished from conventional trapped modes occurring outside of these frequency intervals, wherein the system is effectively closed.

Since BIC do not couple with outgoing waves, their quality factor is infinite. Furthermore,  they cannot be observed in the far field. Conversely, time-reversal symmetry implies that BIC also do not couple with incoming waves and so cannot be excited by incident radiation. BIC only exist for idealised systems and under contrived conditions; small perturbations, say involving geometry, material properties, dissipation or nonlinearity, almost always lead to coupling with incoming and outgoing waves. 
BIC then become leaky states with high quality factors (i.e., long life times relative to the period of oscillation) that can be resonantly excited by incident radiation, albeit slowly, giving rise to narrowband resonances that can be observed in the far-field and exhibit significant field enhancements in the near-field. Leaky states derived from BIC are called asymptotic- or quasi-BIC, indicating their asymptotic approach to BIC as some perturbation parameter, say $\epsilon$, vanishes. Notably, the latter constitutes a singular asymptotic limit, in which, for example, the quality factor diverges. To date, the asymptotic approach of quasi-BIC to BIC has been theoretically studied mainly in the context of embedded guided modes in periodic structures \cite{Bonnet:94,Venakides:03,Shipman:05,Porter:05,Yoon:15,Monticone:17,Bulgakov:17,Koshelev:18,Wu:19,Bykov:19,Overvig:20}, in which case perturbation theory \cite{Yuan:18,Yuan:20,Yuan:20b,Hu:20,Hu:20b} reveals algebraically singular quality factors scaling like $1/\epsilon^s$, where $s$ is a positive integer (typically $s=2,4$ or $6$) and $\epsilon$ represents the perturbation of the Bloch wavenumber from its value at which an embedded guided mode exists. 

\begin{figure}[t!]
\begin{center}
\includegraphics[scale=0.45]{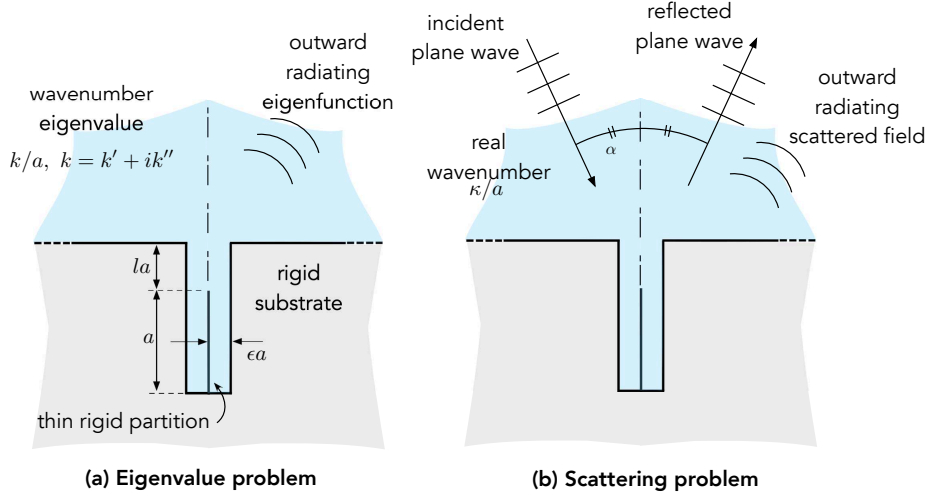}
\caption{(a) Eigenvalue problem governing the acoustic leaky states of a partially  partitioned slit channel connected to a half plane. (b) Associated scattering problem with an incident plane wave.}
\label{fig:dimsketch}
\end{center}
\end{figure}
In this paper we use singular perturbation methods \cite{Hinch:book} to asymptotically study a simple setup that supports quasi-BIC whose quality factor diverge \emph{exponentially} as a perturbation parameter $\epsilon$ vanishes. We also study the exponentially resonant excitation of these states by incident radiation. The setup is shown in Fig.~\ref{fig:dimsketch}. We consider planar acoustic waves in the domain exterior to a flat rigid substrate into which a perpendicular rectangular slit is embedded. Attached to the bottom of the slit is a rigid partition dividing a lower part of the slit into two channels of equal width. We denote the partition length by $a$, the slit length by $(1+l)a$, where $l>0$ is a parameter, and the slit width by $2\epsilon a$. The thickness of the partition is neglected. The leaky states, i.e., quasi-normal modes \cite{Ching:98}, of this configuration are governed by an eigenvalue problem where there is no incident field and the eigenfunction satisfies an outward-radiation condition at large distances (Fig.~\ref{fig:dimsketch}a). The wavenumber, denoted $k/a$ so that $k$ is dimensionless, serves as the eigenvalue. Since the system is open, $k$ is in general complex valued. Writing $k=k'+ik''$, where $k'$ and $k''$ are real, the quality factor is asymptotically $Q=k'/|2k''|$ for small $k''/k'$ \cite{Collin:07}. The excitation of leaky modes is studied by considering an associated scattering problem where the wavenumber, denoted $\kappa/a$, is real and there is an incident field, which we take to be a plane wave incoming at an angle $\alpha$ to the partition (Fig.~\ref{fig:dimsketch}b). 

\begin{figure}[b!]
\begin{center}
\includegraphics[scale=0.45]{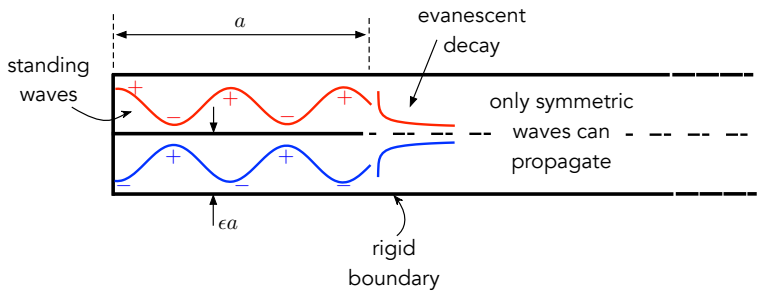}
\caption{A semi-infinite slit channel with a finite partition supports anti-symmetric bound states in the continuum (BIC). The anti-symmetric leaky states of the setup in Fig.~\ref{fig:dimsketch}a that approach the BIC of the semi-infinite geometry as $\epsilon\to0$ are called quasi-BIC.}
\label{fig:boundsketch}
\end{center}
\end{figure}
To see why this setup supports quasi-BIC, consider the semi-infinite geometry shown in Fig.~\ref{fig:boundsketch}. This geometry is just the partitioned slit shown in Fig.~\ref{fig:dimsketch}, infinitely extended away from the partition rather than connecting to a half-plane. A more relevant perspective is that the finite slit domain degenerates to the semi-infinite one in the limit $\epsilon\to0$ if the position along the slit is held fixed. The degenerate geometry is a special case of the more general obstacle-in-channel class of configurations that have been extensively searched for BIC in acoustics, water waves and photonics  \cite{Parker:66,Parker:67,Cumpsty:71,Koch:83,Evans:93,Evans:94,Mciver:98,Linton:98,Groves:98,Davies:98,Linton:02,Linton:07,Duan:07,Cobelli:11,Hsu:16,Sargent:19}. In particular, the semi-infinite geometry shown in Fig.~\ref{fig:boundsketch} is known to support BIC due to symmetry. Indeed, in light of the domain's symmetry, eigenfunctions are either symmetric or anti-symmetric with respect to the partition. For $k<\pi/2\epsilon$, however, only the fundamental, transversally uniform and hence symmetric, waveguide mode is propagating in the unified part of the channel. Accordingly, the eigenvalue problem restricted to anti-symmetric eigenfunctions in this wavenumber regime effectively corresponds to a closed system, in which case discretely distributed real-valued eigenvalues are to be expected. As depicted in Fig.~\ref{fig:boundsketch}, the BIC consist of standing waves at dimensionless wavenumbers $k\approx (1/2+n)\pi$, corresponding to an integer $n+1$ number of quarter-wavelengths fitted along the partition, with order $\epsilon$ corrections associated with end effects at the edge of the partition and with the eigenfunctions decaying exponentially on a length scale $\epsilon a$ away from the partition. A rigorous and constructive proof of existence employing a modified residue-calculus method was provided by Evans, Linton and Ursell \cite{Evans:93} (for a more general class of geometries including the case of an off-centred thin partition, which remarkably can also support BIC). The same paper also presents an intuitive derivation of accurate approximate formulae to be discussed later. Returning to the finite-slit geometry of Fig.~\ref{fig:dimsketch}, it is plausible that the same picture applies, only that now propagating waves are `turned-on' at the aperture by the eigenfunction's evanescent tail, which has decayed exponentially a distance $la$ at a rate of order $1/\epsilon a$. This, in turn, suggests leakage to infinity and corrections to the standing waves in the vicinity of the partition that are exponentially small in $\epsilon/l$. 

Besides being a case study for the asymptotic properties of quasi-BIC, the present study may also be relevant to `metamaterial' modelling and design, where cavities and channels supporting resonances are commonly used in order to manipulate waves; in particular, substrates and plates featuring narrow slits have been extensively used to study the phenomenon of extraordinary transmission \cite{Porto:99,Takakura:01,Garcia:03,Bravo:04,Suckling:04,Christensen:08,Ward:15,Moleron:16}. These studies exploit the conventional standing-wave resonances of a slit cavity, which are associated with cylindrical monopole-like radiation emitted from the slit aperture and quality factors and field enhancements that are algebraically large in the cavity aspect ratio. The asymptotic properties of such resonances have been studied extensively using various methods \cite{Joly:06,Joly:06b,Joly:08,Lin:17,Evans:18,Holley:19,Brandao:20,Chesnel:21,Chesnel:21b}. In this context, the present work shows that adding a partition to a slit introduces an extra set of resonances, which lie in the same frequency regime but are distinct in that they exhibit exponentially large quality factors and field enhancements and that cylindrical dipole-like radiation is emitted from the slit aperture. This added freedom could be used to explore novel BIC-based metamaterial phenomena \cite{Hsu:16,Koshelev:19}, in particular using the analytical results developed herein.   

Building on previous singular-perturbation analyses of resonant phenomena involving narrow slits \cite{Schnitzer:17,Evans:18,Holley:19,Brandao:20,Chesnel:21,Chesnel:21}, we shall use the method of matched asymptotic expansions \cite{Hinch:book} to construct approximations for the wave field by systematically relating asymptotic expansions whose domain of validity spatially overlap. Furthermore, as in \cite{Schnitzer:17,Evans:18,Holley:19}, we shall use conformal mappings \cite{Brown:09} to solve certain `inner' problems that arise as part of the analysis, in this case associated with regions near the partition edge and slit aperture. The present problem differs from those previous singular-perturbation analyses (as well as other asymptotic approaches to slit resonances \cite{Lin:17} and quasi-BIC for periodic media \cite{Yuan:18,Yuan:20,Yuan:20b,Hu:20,Hu:20b}), however, in that exponentially small terms must be calculated. In particular, $k''$ is exponentially small and beyond all orders of the expansion for $k$ in powers of $\epsilon$. Nonetheless, we shall see that a leading approximation for $k''$ can be derived based on knowledge of the leading terms in the expansions for $k$ and the eigenfunction, though including regions away from the partition where the latter is exponentially small. This is done using a reciprocal relation which represents a balance between the rate of change of energy of the state and the energy loss by radiation to infinity. The idea of isolating exponentially small terms using appropriate integral balances is commonly used in scenarios involving exponentially weak radiation losses  \cite{Segur:12}. In the context of the scattering problem, we use a similar approach to calculate the exponential asymptotics describing the unconventional resonant excitation of the quasi-BIC based on the asymptotic approximations derived for the eigenvalue problem. 

The paper is structured as follows. In \S\ref{sec:eprob}, we formulate the eigenvalue problem governing the leaky states of the configuration (Fig.~\ref{fig:dimsketch}a). In \S\ref{sec:bound}, we analyse the anti-symmetric states of the configuration in the limit $\epsilon\to0$. Up to exponentially small terms, they are localised to the vicinity of the partition, where they appear identical to the BIC of the semi-infinite configuration (Fig.~\ref{fig:boundsketch}). The analysis in this section describes the eigenfunctions in those regions and furnishes a real-valued expansion for $k$ (we suffice with calculating the first two terms). In \S\ref{sec:qbound}, we extend the asymptotic description to regions away from the partition, where the eigenfunctions are exponentially smaller in magnitude. We then use the reciprocal relation from \S\ref{sec:eprob} to calculate an exponentially small leading approximation for $k''$. In \S\ref{sec:scat}, we formulate the scattering problem (Fig.~\ref{fig:dimsketch}b) and derive an associated reciprocal relation. We then use the eigenfunction asymptotics and the latter reciprocal relation to find asymptotic approximations describing the exponentially resonant excitation of the quasi-BIC. We give concluding remarks in \S\ref{sec:conc}. 

Throughout the paper, we will compare our asymptotic formulae with numerical results. These were obtained by numerically solving both the eigenvalue problem formulated in \S\ref{sec:eprob} and the scattering problem formulated in \S\ref{sec:scat} based on the formulation and approximation of integral equations which arise from representing solutions using eigenfunction expansions and Fourier transforms in finite and semi-infinite domains, respectively. Details of the numerical schemes are provided as supplementary material \footnote{The supplementary material describes the semi-analytical schemes used to solve the eigenvalue problem formulated in \S\ref{sec:eprob} and the scattering problem formulated in \S\ref{sec:scat} numerically, for the purpose of comparison with the asymptotic approximations derived herein.}.

\begin{figure}[b!]
\begin{center}
\includegraphics[scale=0.65]{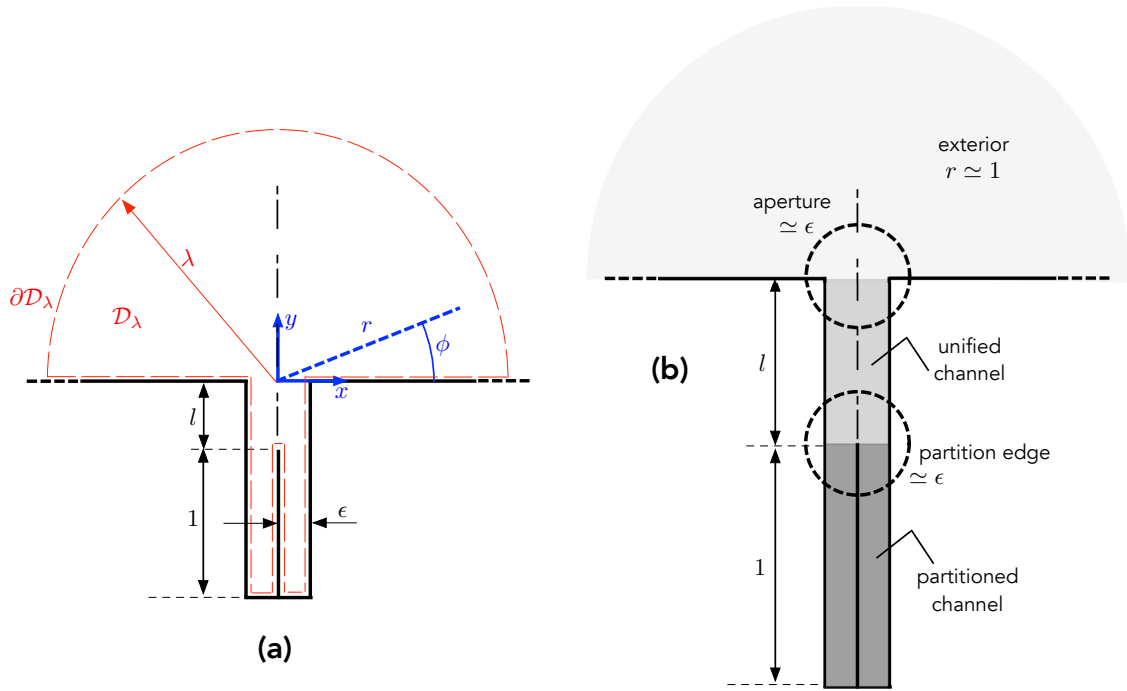}
\caption{(a) Dimensionless sketch of the geometry. (b) Asymptotic regions in the limit $\epsilon\ll1$.}
\label{fig:nondim}
%\label{fig:nondim}a
%\label{fig:nondim}b
\end{center}
\end{figure}
\section{Eigenvalue problem}\label{sec:eprob}
\subsection{Geometry and problem formulation}
It is convenient to normalise lengths by $a$ and work with the rescaled wavenumber eigenvalue $k/a$, where $k=k'+ik''$ as in the introduction. Fig.~\ref{fig:nondim}a shows a dimensionless sketch of the geometry and defines Cartesian coordinates $(x,y)$ and polar coordinates $(r,\phi)$. We assume the convention where the wave field varies with time $t$ like $\exp(-ikt/a)$. In the absence of any source of energy, an outward-radiating wave field cannot grow in time; thus $k''\le0$. The reduced wave field, or eigenfunction, obtained by suppressing the time dependence is denoted by $u$. It is determined up to a multiplicative constant and taken to be dimensionless without loss of generality. 

With these conventions, the eigenvalue problem consists of the Helmholtz equation
\begin{equation}\label{Helm}
\nabla^2u+k^2u=0,
\end{equation}
the Neumann boundary condition
\begin{equation}\label{Nbc}
\pd{u}{n}=0
\end{equation}
and an outward-radiation condition as $r\to\infty$. The latter condition is implemented by requiring that the eigenfunction's far field can be expanded using Hankel functions of the first kind.

In the next two sections, we study the above eigenvalue problem in the limit $\epsilon\ll1$. Specifically, our interest is in the family of quasi-BIC discussed in the introduction, for which the eigenfunctions are anti-symmetric, i.e., odd in $x$, and for which $k\simeq 1$, where $\simeq$ hereafter means `of asymptotic order'. The latter condition implies that, for small $\epsilon$, anti-symmetric waves are evanescent in the unified portion of the slit channel. Given the anti-symmetry of the eigenfunctions, it will sometimes be convenient to consider one half of the fluid domain ($x>0$, or $x<0$) using the Dirichlet condition
\begin{equation}\label{Dir}
u=0 \quad \text{at} \quad x=0 \quad (y>-l).
\end{equation}
%For $\epsilon\ll1$, the discussion in the introduction anticipates anti-symmetric states that are nearly bound to the partition, with only very weak leakage to infinity via the slit opening. 

\subsection{Reciprocal relation}\label{ssec:recip}
Critical to our analysis is an exact reciprocal relation obtained by applying Green's second identity to the field $u$ and its complex conjugate $u^*$, 
\begin{equation}\label{identity1}
\oint_{\partial \mathcal{D}_{\lambda}\!\!}\left(u^*\pd{u}{n}-u\pd{u^*}{n}\right)\,dl+4ik'k''\int_{\mathcal{D}_{\lambda}}|u|^2\,dA=0,
\end{equation}
where $dl$ and $dA$ are infinitesimal length and area elements, respectively, and we choose the domain $D_{\lambda}$ as the  union of the slit region and the subset $r<\lambda$ of the exterior half-plane, for some $\lambda>0$ (see Fig.~\ref{fig:nondim}a). With that choice, and using the Neumann condition \eqref{Nbc}, identity \eqref{identity1} can be written 
\begin{equation}\label{identity2}
k''=\frac{i}{4k'}\frac{\lambda\int_0^\pi\left(u^*\pd{u}{r}-u\pd{u^*}{r}\right)_{r=\lambda}\,d\phi}{\int_{\mathcal{D}_\lambda}|u|^2\,dA}.
\end{equation}
Note that $k''$ cannot depend upon $\lambda$, which is arbitrary.  This identity represents a balance between the time rate of change of the state's energy, which is proportional to $k''$, and the energy lost by radiation to infinity. It will be used in \S\ref{ssec:calckpp} to calculate an asymptotic approximation for $k''$. 

\section{Apparent bound states in the continuum} \label{sec:bound}
Fig.~\ref{fig:nondim}b portrays the asymptotic regions which are relevant to the analysis of the anti-symmetric states in the limit $\epsilon\ll1$ . As we shall see, the eigenfunctions are strongly localised to the partition-channel and partition-edge regions. In fact, our analysis in the present section will confirm that, up to exponentially small orders, the eigenfunctions involve only those regions and are therefore apparently equivalent to the BIC of the semi-infinite configuration discussed in the introduction. The unified-channel, aperture and exterior regions, where the eigenfunctions are exponentially smaller than in the partitioned-channel and partition-edge region, will be considered in \S\ref{sec:qbound}. %, with the purpose of calculating a leading approximation for $k''$, which is beyond all orders of the expansion for $k$. 

\subsection{Partitioned channel}\label{ssec:part}
Without loss of generality, let the eigenfunction $u$ be of order unity in the partitioned-channel region. To analyse that region, we define the stretched coordinates $x=\epsilon X$ and define $u(x,y;\epsilon)=U(X,y;\epsilon)$, with $-1-l<y<-l$ and, focusing on the right-hand side of the partitioned-channel region, $0<X<1$. The eigenfunction $U$ satisfies
\begin{equation}\label{helm part}
\pd{^2U}{X^2}+\epsilon^2\left(\pd{U}{y^2}+k^2U\right)=0,
\end{equation}
the Neumann boundary condition 
\begin{equation}\label{N part a}
\pd{U}{X}=0 \quad \text{at} \quad X=0,1
\end{equation}
and, naively, the Neumann boundary condition
\begin{equation}\label{N part b}
\pd{U}{y}=0 \quad \text{at} \quad y=-1-l.
\end{equation}
Strictly speaking, the latter condition should be verified by matching to an `end' region $y+1+l\simeq\epsilon$. Nonetheless, we shall assume that \eqref{N part b} holds at least up to exponentially small orders for reasons noted at the end of this section. Additionally, conditions at $y\to -l$ are to be derived by asymptotic matching with the partition-edge region. 
%\begin{figure}[t!]
%\begin{center}
%\includegraphics[scale=0.6]{regions.eps}
%\caption{Asymptotic regions in the limit $\epsilon\ll1$.}
%\label{fig:nondim}b
%\end{center}
%\end{figure}

The wavenumber is expanded as  
\begin{equation}
k=k_0+\epsilon k_1+\cdots \quad \text{as} \quad \epsilon\to0.
\end{equation}
Similarly, the eigenfunction is expanded as 
\begin{equation}\label{part expansion}
U(X,y;\epsilon)=U_0(X,y)+\epsilon U_1(X,y)+\epsilon^2U_2(X,y)+\epsilon^3U_3(X,y)+ \cdots \quad \text{as} \quad \epsilon\to0.
\end{equation}
At orders unity and $\epsilon$, \eqref{helm part} gives
\begin{equation}
\pd{^2U_0}{X^2}=0, \quad \pd{^2U_1}{X^2}=0.
\end{equation}
Together with the respective orders of the Neumann conditions \eqref{N part a}, we find 
\refstepcounter{equation}
$$
U_0(X,y)=U_0(y), \quad U_1(X,y)=U_1(y). 
\eqno{(\theequation{\mathrm{a},\!\mathrm{b}})}
$$
At order $\epsilon^2$, \eqref{helm part} gives
\begin{equation}\label{U0 initial eq}
\pd{^2U_2}{X^2}+\frac{d^2U_0}{dy^2}+k_0^2U_0=0.
\end{equation}
Integration with respect to $X$ and using the Neumann conditions \eqref{N part a} at the respective order, we find the one-dimensional Helmholtz equation 
\begin{equation}\label{U0 eq}
\frac{d^2U_0}{dy^2}+k_0^2U_0=0.
\end{equation}
One boundary condition for \eqref{U0 eq} is obtained from the leading-order balance of the Neumann condition \eqref{N part b},
\begin{equation}\label{U0 Nbc}
\frac{dU_0}{dy}=0 \quad \text{at} \quad y=-1-l. 
\end{equation}
A second boundary condition is obtained by anticipating that the eigenfunction is asymptotically small in the partition-edge region. (As noted in \S\ref{ssec:edge}, assuming otherwise leads to a  contradiction in the partition-edge region.) That implies the matching condition 
\begin{equation}\label{U0 res cond}
U_0=0 \quad \text{at} \quad y=-l. 
\end{equation}

Solving the eigenvalue problem \eqref{U0 eq}--\eqref{U0 res cond} gives the leading-order eigenvalues,
\begin{equation}\label{k0 sol}
k_0=\left(\frac{1}{2}+n\right)\pi, \quad n=0,1,2,\ldots,
\end{equation}
and corresponding eigenfunctions
\begin{equation}\label{U0 sol}
U_0=\cos(k_0(y+1+l)), 
\end{equation}
where without loss of generality we choose the constant of proportionality to be unity. For simplicity we keep the mode number $n$ implicit in our notation for the wavenumber eigenvalue and eigenfunction. 

For later reference, we note the behaviour
\begin{equation}\label{U0 behaviour}
U_0\sim -\sigma_n k_0(y+l) \quad \text{as} \quad y\to -l,
\end{equation}
where we define $\sigma_n=(-1)^n$. We also note that substitution of the one-dimensional Helmholtz equation \eqref{U0 eq} into \eqref{U0 initial eq} and using the order $\epsilon^2$ balance of the Neumann conditions \eqref{N part a} shows that $U_2(X,y)=U_2(y)$. 

\subsection{Partition edge}\label{ssec:edge}
\begin{figure}[t!]
\begin{center}
\includegraphics[scale=0.7]{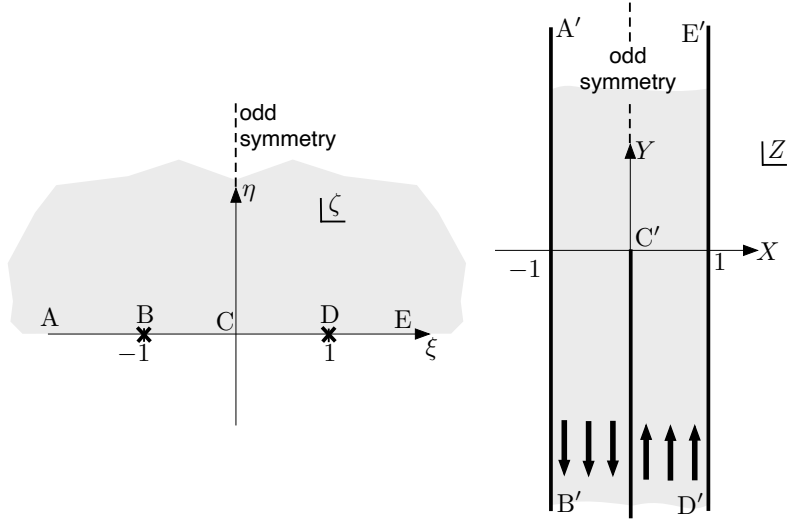}
\caption{Mapping from the upper half of an auxiliary complex $\zeta$ plane to the domain in the complex $Z$ plane corresponding to the partition-edge region.}
\label{fig:edge}
\end{center}
\end{figure}

To analyse the partition-edge region we employ the strained coordinates $X=x/\epsilon$ and $Y=(y+l)/\epsilon$ and define $u(x,y;\epsilon)=V(X,Y;\epsilon)$, with $(X,Y)$ in the domain shown on the right-hand side of Fig.~\ref{fig:edge}. The eigenfunction $V$ satisfies 
\begin{equation}\label{V eq}
\pd{^2V}{X^2}+\pd{^2V}{Y^2}+\epsilon^2k^2V=0,
\end{equation}
the Neumann boundary condition
\begin{equation}\label{V N}
\pd{V}{N}=0,
\end{equation}
where $\partial/\partial{N}=\bn\bcdot(\be_x\partial/\partial{X}+\be_y\partial/\partial{Y})$, and conditions to be derived by matching with the partitioned- and unified-channel regions. In particular, we anticipate that $V$, up to exponential orders in $\epsilon$, decays exponentially as $Y\to+\infty$. We also enforce that $V$ is odd in $X$. 

The behaviour \eqref{U0 behaviour} suggests $V\simeq \epsilon$. Accordingly, $V$ is expanded as
\begin{equation}\label{edge expansion}
V(X,Y;\epsilon)= \epsilon V_0(X,Y) + \cdots \quad \text{as} \quad \epsilon\to0.
\end{equation}
From \eqref{V eq} and \eqref{V N}, $V_0$ satisfies Laplace's equation
\begin{equation}
\pd{^2V_0}{X^2}+\pd{^2V_0}{Y^2}=0
\end{equation}
and the Neumann condition
\begin{equation}
\pd{V_0}{N}=0,
\end{equation}
respectively. The anticipated decay towards the unified-channel region implies
\begin{equation}\label{V1 decay}
V_0\to0\quad \text{as} \quad Y\to+\infty \quad (|X|<1),
\end{equation}
while matching the partitioned-channel expansion \eqref{part expansion} to order unity with the partition-edge expansion \eqref{edge expansion} to order $\epsilon$, using \eqref{U0 behaviour}, yields the condition
\begin{equation}\label{V1 matching}
V_0\sim -\sigma_n k_0 Y \quad \text{as} \quad Y\to-\infty \quad (0<X<1).
\end{equation}
The corresponding condition for $-1<X<0$ follows from $V_0$ being odd in $X$.

To solve the above potential problem, we introduce the complex variable $Z=X+iY$, the auxiliary complex variable $\zeta=\xi+i\eta$ and the mapping \cite{Brown:09}
\begin{equation}
Z=1+\frac{i}{\pi}\left[\log (\zeta-1)+\log(\zeta+1)\right]
\end{equation}
depicted in Fig.~\ref{fig:edge} 
from the upper half of the $\zeta$ plane to the fluid domain of the partition-edge region. It is useful to derive the asymptotic behaviours of the mapping as $\zeta$ approaches infinity as well as the singular points $\zeta=\pm1$. As $\zeta\to\infty$ in the upper half-plane, we find
\refstepcounter{equation}
$$
\label{edge mapping inf}
X= 1-\frac{2\theta}{\pi}+\mathcal{O}\left(\frac{1}{\zeta^2}\right), \quad Y= \frac{2\ln|\zeta|}{\pi} +\mathcal{O}\left(\frac{1}{\zeta^2}\right), 
\eqno{(\theequation{\mathrm{a},\!\mathrm{b}})}
$$
where $\theta=\arg(\zeta)$ with $0\le\theta\le\pi$. As $\zeta\to1$ in the upper half-plane, we find
\refstepcounter{equation}
$$
\label{edge mapping near 1}
X=1-\frac{\varphi}{\pi}+\mathcal{O}(\zeta-1), \quad Y=\frac{1}{\pi}\ln|\zeta-1|+\frac{\ln 2}{\pi} + \mathcal{O}(\zeta-1),
\eqno{(\theequation{\mathrm{a},\!\mathrm{b}})}
$$
where $\varphi=\arg(\zeta-1)$ with $0\le\varphi\le\pi$. The corresponding behaviour as $\zeta\to-1$ readily follows from the symmetry of the mapping about the $\eta$ axis, with $Z(\pm\xi,\eta)=\pm X+iY$. 

Let $V_0(X,Y)=v(\xi,\eta)$. The harmonic function $v(\xi,\eta)$ is anti-symmetric about the $\eta$ axis, decays as $|\zeta|\to\infty$ in the upper half-plane and satisfies a Neumann condition on $\eta=0$ except at the singular points $\zeta=\pm1$, where \eqref{V1 matching} and \eqref{edge mapping near 1} together give 
\begin{equation}
v\sim \mp\frac{\sigma_n k_0}{\pi}\ln|\zeta\mp1| \quad \text{as} \quad \zeta\to\pm1.
\end{equation}
The latter condition, which effectively prescribes opposite monopoles at $\zeta=\pm1$, can also be derived by matching fluxes between the partitioned-channel partition-edge regions. The solution to the problem governing $v$ is readily seen to be 
\begin{equation}\label{v sol}
v=-\frac{\sigma_n k_0}{\pi}\left(\ln|\zeta-1|-\ln|\zeta+1|\right).
\end{equation}

The limit 
\begin{equation}\label{beta def}
\beta=\lim_{Y\to-\infty}\left\{V_0+\sigma_n k_0Y\right\},
\end{equation}
taken in $0<X<1$, will be required at the next order of the analysis. Using \eqref{edge mapping near 1} and \eqref{v sol}, we find 
\begin{equation}
\beta=\frac{\sigma_n k_0\ln 4}{\pi}.
\end{equation}

Another output of the solution \eqref{v sol} is the behaviour
\begin{equation}\label{exp behaviour}
V_0\sim \frac{2\sigma_n k_0}{\pi}\sin\frac{\pi X}{2}\exp\left(-\frac{\pi Y}{2}\right) \quad \text{as} \quad Y\to+\infty \quad (|X|<1),
\end{equation}
which is derived using \eqref{edge mapping inf}. We will use this behaviour in \S\ref{sec:qbound} to match with the unified-channel region. For the time being, it serves to confirm that the eigenfunction is exponentially small in the unified-channel, aperture and exterior regions. 

We remark in passing that if we had not assumed in \S\ref{ssec:part} that  $U_0(-l)=0$, then $V$ would be order unity; a leading approximation would satisfy the same problem as $V_0$ above, only that it would require to approach the non-zero constants $\pm U_0(-l)$ as $Y\to-\infty$, respectively in $0<\pm X<1$. The latter problem, however, does not possess a solution which is why we rejected this possibility. 

\subsection{Higher-order analysis}
Returning to the right-hand side of the partitioned-channel region, we find from the order $\epsilon^3$ balance of \eqref{helm part} that $U_1$ satisfies
\begin{equation}\label{U1 eq initial}
\frac{d^2U_1}{dy^2}+k_0^2U_1+\pd{^2U_3}{X^2}=-2k_0k_1U_0.
\end{equation}
Integration with respect to $X$, together with the respective order of the Neumann boundary conditions \eqref{N part a}, we find the inhomogeneous one-dimensional Helmholtz equation
\begin{equation}\label{U1 eq}
\frac{d^2U_1}{dy^2}+k_0^2U_1=-2k_0k_1U_0.
\end{equation}
This equation is supplemented by the Neumann condition
\begin{equation}\label{U1 bc n}
\frac{dU_1}{dy}=0 \quad \text{at} \quad y=-1-l,
\end{equation}
which follows from the order $\epsilon$ balance of the Neumann condition \eqref{N part b}, and the condition
\begin{equation}\label{U1 bc beta}
U_1 = \beta  \quad \text{at} \quad y=-l,
\end{equation}
which follows from matching the partition-edge expansion \eqref{edge expansion} and the partitioned-channel expansion \eqref{part expansion} to order $\epsilon$, using \eqref{beta def}.  

The problem for $U_1$ is the same as the singular  homogeneous problem for $U_0$, with the addition of forcing terms. A solution to the former problem therefore exists only under special conditions on those forcing terms. Subtracting the complex conjugate of \eqref{U0 eq} multiplied by $U_1$ from \eqref{U1 eq} multiplied by $U_0^*$, where an asterisk denotes complex conjugation, and using integration by parts and the boundary conditions \eqref{U0 Nbc}, \eqref{U0 res cond}, \eqref{U1 bc n} and \eqref{U1 bc beta}, we find the solvability condition 
\begin{equation}
\beta \left.\frac{dU_0^*}{dy}\right|_{y=-l}=2k_0k_1\int_{-1-l}^{-l}|U_0|^2\,dy.
\end{equation}
Substituting \eqref{k0 sol} and \eqref{U0 sol}, we find the first wavenumber correction
\begin{equation}\label{k1 sol}
k_1 = -k_0\frac{\ln 4}{\pi}.
\end{equation}
\begin{figure}[t!]
\begin{center}
\includegraphics[scale=0.5]{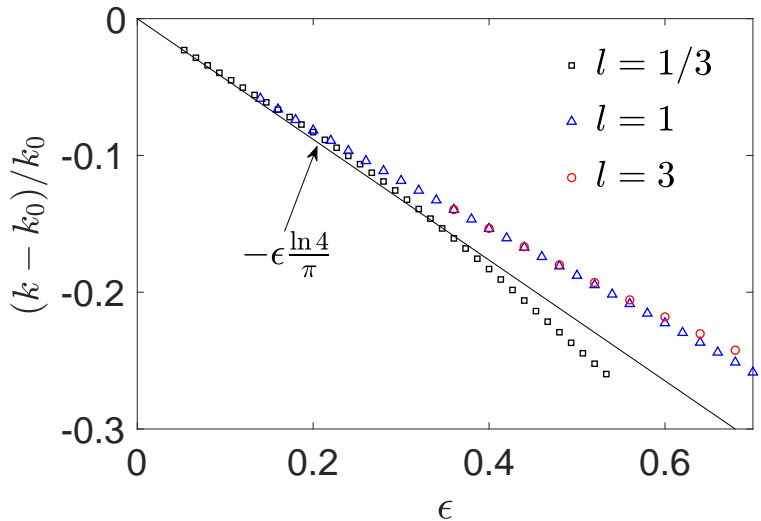}
\caption{The relative shift of the wavenumber $k$ from its leading approximation $k_0$ as a function of $\epsilon$, for the fundamental anti-symmetric mode $n=0$ for which $k_0=\pi/2$. Solid curve: asymptotic approximation \eqref{kmk0}. Symbols: numerical results for the indicated values of $l$.}
\label{fig:krel}
\end{center}
\end{figure}

Combining \eqref{k0 sol} and \eqref{k1 sol}, the approximation for the wavenumber can be written
\begin{equation}\label{kmk0}
\frac{k-k_0}{k_0}\sim -\epsilon\frac{\ln 4}{\pi} \quad \text{as} \quad \epsilon\to0.
\end{equation}
This approximation is compared with numerical results in Fig.~\ref{fig:krel}. %our asymptotic analysis was carried out in the limit $\epsilon\to0$ with $l$ fixed; it is clear from the preceding analysis that the resulting approximations do not hold for $l\simeq \epsilon$, in which case the partitioned-channel and partition-edge region combine and compressibility effects become important in the latter, and for $(1-l)\simeq\epsilon$, where the partition-edge, unified-channel and aperture regions join and the analogy with the embedded bound states of a semi-infinite channel with a partition is lost. The latter limitation will be more apparent from the analysis in the next section. 
We also note that \eqref{kmk0} is asymptotically consistent with the higher-order approximations in implicit form obtained by Evans, Linton and Ursell \cite{Evans:93} for the BIC of the semi-infinite slit geometry seen in Fig.~\ref{fig:boundsketch}. We note that the approximations in \cite{Evans:93} are derived intuitively by alluding to exact Wiener-Hopf solutions to the Helmholtz equation for a semi-infinite geometry corresponding to our partition-edge region. 

There is no conceptual difficulty in extending the analysis of the present section, involving solely the partitioned-channel and partition-edge regions, to higher orders. In particular, it is clear that to arbitrary power of $\epsilon$ the partitioned-channel eigenfunction remains transversally uniform (thus justifying the use of the Neumann boundary condition \eqref{N part b}) and that the partition-edge eigenfunction decays exponentially, though with compressibility effects entering at order $\epsilon^3$ as a perturbation. Working through these details would furnish an explicit, real-valued, infinite asymptotic expansion in powers of $\epsilon$ for $k$. Instead, we next derive a leading-order approximation for $k''$ which is beyond all orders of the expansion for $k$.

\section{Quasi-bound states in the continuum}\label{sec:qbound}
\subsection{Unified channel}
In this section we extend the asymptotic description of the anti-symmetric modes into the unified-channel and exterior regions, where the eigenfunction exponentially small. Together with the reciprocal identity \eqref{identity2}, this will allow us to calculate a leading approximation for $k''$. We start with the unified-channel region, where we employ the coordinates $(X,y)$, with $-l<y<0$ and $|X|<1$, and define $u(x,y;\epsilon)=W(X,y;\epsilon)$. The eigenfunction $W$ satisfies the  equation 
\begin{equation}\label{W eq}
\pd{^2W}{X^2}+\epsilon^2\pd{^2W}{y^2}+\epsilon^2k^2W=0,
\end{equation}
the Neumann conditions
\begin{equation}\label{W bc}
\pd{W}{X}=0 \quad \text{at} \quad X=\pm1 
\end{equation}
and conditions as $y\to-l$ and $y\to0$ to be derived from asymptotic matching with the partition-edge and aperture  regions, respectively. We also enforce that $W$ is odd in $X$.

Like the partitioned-channel region, the unified-channel region corresponds to a channel of width $\simeq \epsilon$ and length $\simeq 1$. In contrast to the partitioned-channel region, however, the absence of the partition in the unified-channel together with the anti-symmetry of the eigenfunction precludes wave propagation. Rather, the eigenfunction in the unified-channel region is, to leading-order, non-uniform in the transverse direction, attenuating exponentially away from the partition edge. An asymptotic expansion in this region can be derived using a WKB ansatz \cite{Hinch:book}; alternatively, it can be extracted from an exact separation-of-variables series solution to the Helmholtz equation \eqref{W eq} in the rectangular domain corresponding to the unified-channel region. Given \eqref{exp behaviour} and the need to match with the partition-edge region, we find the leading-order approximation
\begin{equation}\label{unified expansion}
W(X,y;\epsilon)= \epsilon  \frac{2\sigma_n k_0}{\pi} \sin \frac{\pi X}{2} \exp\left\{-\frac{\pi}{2\epsilon}(y+l)\right\} +\cdots \quad \text{as} \quad \epsilon\to0.
\end{equation}
We note that the above approximation is harmonic even though the unified-channel region is not subwavelength.  

\subsection{Aperture}
\begin{figure}[t!]
\begin{center}
\includegraphics[scale=0.7]{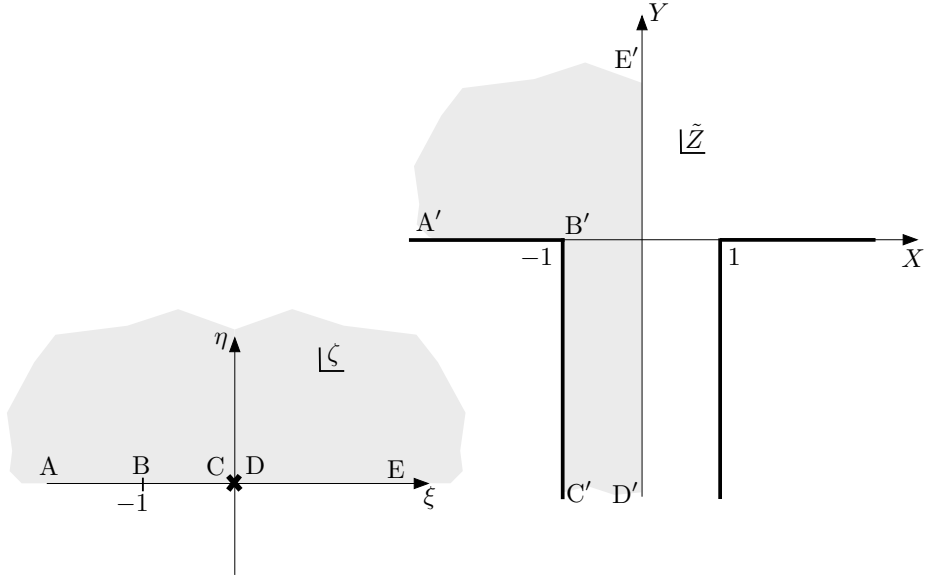}
\caption{Mapping from the upper half of an auxiliary complex $\zeta$ plane to the domain in the complex $\tilde{Z}$ plane corresponding to the left-hand side of the aperture region.}
\label{fig:opening}
\end{center}
\end{figure}

We next consider the aperture region bridging the unified-channel and exterior regions. To study this region, we employ the strained coordinates $X$ and $\tilde{Y}=y/\epsilon$ and define $u(x,y;\epsilon)=S(X,\tilde{Y};\epsilon)$. The domain of $X$ and $\tilde{Y}$ is as shown on the right-hand side of Fig.~\ref{fig:opening}. In that domain, the eigenfunction $S$ satisfies the Helmholtz equation
\begin{equation}
\pd{^2S}{X^2}+\pd{^2S}{\tilde{Y}^2}+\epsilon^2k^2S=0,
\end{equation}
the Neumann condition
\begin{equation}
\pd{S}{N}=0
\end{equation}
and far-field conditions to be derived by matching with the unified-channel and exterior regions. We also enforce that $S$ is odd in $X$. 

The form of the unified-channel expansion \eqref{unified expansion} implies that $S$ should be expanded  as 
\begin{equation}\label{S exp}
S= \epsilon \exp\left(-\frac{\pi l}{2\epsilon}\right)S_0(X,\tilde{Y}) + \cdots \quad \text{as} \quad \epsilon\to0. 
\end{equation}
From \eqref{Helm}, $S_0$ satisfies Laplace's equation
\begin{equation}
\pd{^2S_0}{X^2}+\pd{^2S_0}{\tilde{Y}^2}=0,
\end{equation}
while \eqref{Nbc} gives the Neumann condition
\begin{equation}\label{S0 NBc}
\pd{S_0}{{N}}=0.
\end{equation}
With regards to matching with the exterior region, the anti-symmetry of the eigenfunction implies the attenuation condition
\begin{equation}\label{S0 decay}
S_0\to0 \quad \text{as} \quad X^2+\tilde{Y}^2\to\infty \quad (\tilde{Y}>0).
\end{equation}
Indeed, the far-field of an anti-symmetric harmonic function in a half-plane domain  either grows algebraically or decays algebraically. The former possibility is non-sensical as it would imply that the exterior eigenfunction is asymptotically larger than the aperture eigenfunction; indeed, only a decaying aperture eigenfunction is consistent with the exterior eigenfunction being outward-radiating.  A last condition on $S_0$ is obtained by  matching the leading orders of the unified-channel expansion \eqref{unified expansion} and the aperture expansion \eqref{S exp}. This gives 
\begin{equation}\label{S0 exp}
S_0 \sim \sigma_n\frac{2k_0}{\pi}\sin \frac{\pi X}{2}\exp\left\{-\frac{\pi}{2}\tilde{Y}\right\} \quad \text{as} \quad \tilde{Y}\to-\infty \quad (|X|<1). 
\end{equation}

To solve for $S_0$ we introduce the complex variable $\tilde{Z}=X+i\tilde{Y}$, the auxiliary variable $\zeta=\xi+i\eta$ and the mapping \cite{Brown:09}
\begin{equation}
\tilde{Z}=\frac{2i}{\pi}(1+\zeta)^{1/2}+\frac{i}{\pi}\log\frac{(1+\zeta)^{1/2}-1}{(1+\zeta)^{1/2}+1}
\end{equation}
from the upper half of the complex $\zeta$ plane to the left-hand side of the aperture domain (see Fig.~\ref{fig:opening}). As $\zeta\to0$ in the upper half-plane,  
\refstepcounter{equation}
$$
\label{X tilde Y 0}
X= -\frac{\theta}{\pi} +o(1), \quad \tilde{Y}=\frac{1}{\pi}\ln |\zeta|+\frac{2}{\pi}\left(1-\ln 2\right) + o(1). 
\eqno{(\theequation{\mathrm{a},\!\mathrm{b}})}
$$
where $\theta=\arg{\zeta}$ with $0\le\theta\le\pi$. 
As $\zeta\to\infty$ in the upper half-plane, 
\begin{equation}\label{opening large zeta}
\tilde{Z}=\frac{2i}{\pi}\zeta^{1/2}+\mathcal{O}\left(\frac{1}{\zeta^{1/2}}\right). 
\end{equation}

Let $S_0(X,\tilde{Y})=s(\xi,\eta)$. The harmonic function $s(\xi,\eta)$ satisfies a decay condition as $|\zeta|\to\infty$ in the upper half-plane, a Neumann condition on the negative $\xi$ axis, a Dirichlet condition on the positive $\xi$ axis and the matching condition
\begin{equation}
s\sim -\frac{4k_0\sigma_n}{\pi e}\frac{\sin(\theta/2)}{|\zeta|^{1/2}} \quad \text{as} \quad \zeta\to0
\end{equation}
in the upper half-plane, which is derived from \eqref{S0 exp} and \eqref{X tilde Y 0}. It is easy to see that the solution is identical to that asymptotic behaviour, i.e., 
\begin{equation}\label{opening solution}
s=-\frac{4 k_0\sigma_n}{\pi e}\frac{\sin(\theta/2)}{|\zeta|^{1/2}}.
\end{equation}
The solution can also be viewed as being proportional to $\mathrm{Im}(1/\zeta^{1/2})$. 

In preparation for matching with the exterior region, we use \eqref{opening large zeta} to extract from the solution \eqref{opening solution} the far-field behaviour
\begin{equation}\label{S0 dipolar behaviour}
S_0 \sim \frac{8 k_0\sigma_n}{\pi^2 e}\frac{\cos\phi}{\tilde{R}} \quad \text{as} \quad \tilde{R}\to\infty,
\end{equation}
where $\tilde{R}^2= X^2+\tilde{Y}^2$ and $\phi$ is the angle measured counter-clockwise from the $X$ direction (i.e., identical to the angle defined in \S\ref{sec:eprob}). The behaviour \eqref{S0 dipolar behaviour} represents a potential dipole.

\subsection{Exterior}
Consider now the exterior region, where the  unstrained coordinates $(x,y)$, alternatively $(r,\phi)$, are held fixed and the eigenfunction is denoted, as in the exact problem formulation, by $u$. The exterior problem consists of the Helmholtz equation \eqref{Helm} in the half-plane $y>0$, with a far-field outward-radiation condition, a Neumann condition at $y=0$ ($x\ne0$) and conditions as $r\to0$ to be derived from matching with the aperture region.

In particular, the aperture expansion \eqref{S exp} together with the dipolar behaviour \eqref{S0 dipolar behaviour} suggests expanding the exterior region as 
\begin{equation}\label{exterior exp}
u(r,\phi;\epsilon) = \epsilon^2 \exp\left(-\frac{\pi l}{2\epsilon}\right)u_0(r,\phi)+\cdots \quad \text{as} \quad \epsilon\to0. 
\end{equation}
From \eqref{Helm}, $u_0$ satisfies the Helmholtz equation 
\begin{equation}
\nabla^2u_0+k_0^2u_0=0,
\end{equation}
while \eqref{Nbc} gives the Neumann condition
\begin{equation}
\pd{u_0}{y}=0 \quad \text{at} \quad y=0 \quad (x\ne0).
\end{equation}
Using \eqref{S0 dipolar behaviour} to match the leading orders of the aperture expansion \eqref{S exp} and exterior expansion \eqref{exterior exp}, we find the singular condition
\begin{equation}
u_0 \sim  \frac{8k_0\sigma_n}{\pi^2e}\frac{\cos\phi}{r} \quad \text{as} \quad r\to0.
\end{equation}
Lastly, we have that $u_0$ satisfies an outward-radiation condition as $r\to\infty$. 

The solution is readily seen to be
\begin{equation}\label{u0 sol}
u_0 = \frac{4ik_0^2\sigma_n}{\pi e} {H}\ub{1}_1(k_0r)\cos\phi,
\end{equation}
where ${H}\ub{1}_1$ denotes the first-order Hankel function of the first kind \cite{Abramowitz:book}. The far-field of $u_0$ can be extracted from \eqref{u0 sol} using the large-argument asymptotic behaviour of the Hankel function \cite{Abramowitz:book}. This gives the dipolar radiation field
\begin{equation}\label{u0 far}
u_0 \sim \frac{2^{1/2}4k_0^{3/2}\sigma_n e^{-i\pi/4}}{\pi^{3/2}e}\frac{\cos\phi}{r^{1/2}}e^{ik_0r} \quad \text{as} \quad r\to\infty. 
\end{equation}

We stress that the exterior expansion is derived in the limit $\epsilon\to0$ with $r$ fixed. It clearly breaks down at large distances $r\simeq 1/\epsilon$, where the wavenumber correction $k_1$ becomes important to leading order. At larger and larger distances higher corrections of the wavenumber become important; eventually, at exponentially large distances $r\simeq 1/|k''|$, the imaginary part of the wavenumber enters the leading-order balance and the eigenfunction begins to grow exponentially, as one would expect for a quasi-normal mode \cite{Ching:98}.  
\begin{figure}[t!]
\begin{center}
\includegraphics[trim=50 0 0 0,scale=0.39]{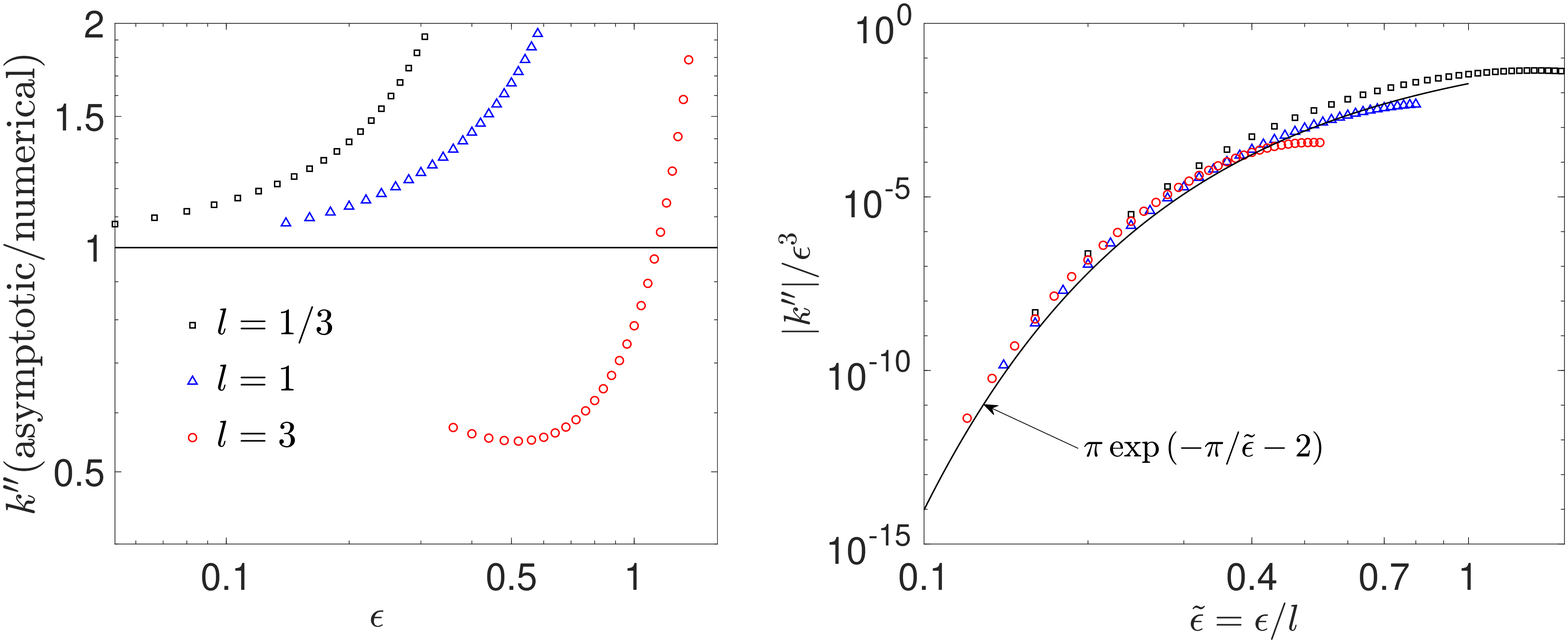}
\caption{The imaginary part $k''$ of the wavenumber $k$ for the fundamental anti-symmetric mode $n=0$. Left: ratio of the asymptotic approximation \eqref{kpp sol} for $k''$ to numerically computed values as a function of $\epsilon$ and for the indicated values of $l$. Right: $k''$ and $\epsilon$ scaled so as to collapse numerical data for different $l$ on a single analytical curve implied by \eqref{kpp sol}.}
\label{fig:kpp}
\end{center}
\end{figure}

\subsection{Imaginary part of the wavenumber}\label{ssec:calckpp}
We are now ready to calculate a leading-order approximation for $k''$. To this end, we employ the reciprocal relation \eqref{identity2} with $1\ll\lambda\ll1/\epsilon$. Then \eqref{exterior exp} together with \eqref{u0 far} gives 
\begin{equation}\label{num as}
\lambda \int_0^\pi\left(u^*\pd{u}{r}-u\pd{u^*}{r}\right)_{r=\lambda}\,d\phi \sim \frac{32ik_0^4}{\pi^2e^2}\epsilon^4\exp\left(-\frac{\pi l}{\epsilon}\right) \quad \text{as} \quad \epsilon\to0.
\end{equation}
The area integral in \eqref{identity2} is dominated by the partitioned-channel region. Thus, using \eqref{part expansion} and \eqref{U0 sol} we find
\begin{equation}\label{denom as}
{\int_{\mathcal{D}_\lambda\!\!}|u|^2\,dA}\sim 2\epsilon\int_{-1-l}^{-l}|U_0|^2\,dy=\epsilon \quad \text{as} \quad \epsilon\to0.
\end{equation}
With \eqref{num as} and \eqref{denom as}, the reciprocal relation \eqref{identity2} yields
\begin{equation}\label{kpp sol}
k''\sim -\frac{8k_0^3\epsilon^3}{e^2\pi^2}\exp\left(-\frac{\pi l}{\epsilon}\right) \quad \text{as} \quad \epsilon\to0.
\end{equation}
As expected, $k''$ is seen to be exponentially small, the order of exponential smallness being proportional to $l/\epsilon$.  In Fig.~\ref{fig:kpp}, the asymptotic approximation \eqref{kpp sol} is compared with numerical results.

\section{Resonant excitation by oblique plane wave}\label{sec:scat}
\subsection{Scattering problem}
In order to illustrate the significance of the quasi-BIC  studied in the preceding sections, we next consider the associated scattering problem described in the Introduction where these states are resonantly excited --- it involves the same geometry and physical assumptions as for the eigenvalue problem formulated in \S\ref{sec:eprob}, only with a real dimensionless wavenumber $\kappa$ and a plane wave at oblique incidence $\alpha$ as shown in Fig.~\ref{fig:dimsketch}b. Our analysis of this scattering problem will focus on the resonant excitation of the quasi-BIC for small $\epsilon$ and will accordingly build on the results already obtained for the eigenvalue problem. %The wave field can be decomposed into a symmetric and anti-symmetric part, our interest being solely in the anti-symmetric part. 

We denote by $v$ the wave field normalised by the amplitude of the incident plane wave. It satisfies the Helmholtz equation 
\begin{equation}\label{Helm S}
\nabla^2v+\kappa^2v=0,
\end{equation}
the Neumann condition 
\begin{equation}\label{Nbc S}
\pd{v}{n}=0
\end{equation}
and an outward radiation condition as $r\to\infty$ in the exterior half-plane, on $v-v_i$, where \begin{equation}\label{incident}
v_i=\exp\{i\kappa(x\sin\alpha-y\cos\alpha)\}
\end{equation}
is the incident plane wave. 

It is convenient to decompose the total field as 
\begin{equation}\label{decompose}
v=v_a+v_s,
\end{equation}
in which
\begin{equation}\label{ambient}
v_a=2\cos(\kappa y \cos\alpha)\exp\left\{i\kappa x \sin\alpha\right\}
\end{equation}
is the ambient field, namely the sum of the incident field $v_i$ and the reflection from the flat substrate in the absence of the slit. The ambient field satisfies the Helmholtz equation \eqref{Helm S} with a Neumann condition on $y=0$. The scattered field $v_s$ satisfies the Helmholtz equation \eqref{Helm S}, a radiation condition as $r\to\infty$ in the exterior half-plane and an inhomogeneous Neumann condition. 
%\begin{equation}\label{Nbc S}
%\pd{v_s}{n}=-\pd{v_a}{n}
%\end{equation}
%and an outward radiation condition as $r\to\infty$. 

The incident field $v_i$ can be decomposed into parts which are symmetric and anti-symmetric with respect to the $y$ axis, respectively. The total field $v$ and ambient field $v_a$ can be similarly decomposed and there is no coupling between the symmetric and anti-symmetric fields. Thus, for normal incidence the fields are perfectly symmetric thence the quasi-BIC, which are anti-symmetric, cannot be excited. We therefore restrict our attention to incidence angles in the range $0<\alpha<\pi/2$. 

\subsection{Reciprocal relation}
We shall employ a reciprocal relation between the scattering problem and the eigenvalue problem of \S\ref{sec:eprob}. To derive it, we apply Green's second identity to $v^*$, the complex conjugate of the total field, and an eigenfunction $u$ over the domain $\mathcal{D}_{\lambda}$ defined in \S\ref{ssec:recip} (Fig.~\ref{fig:nondim}a). Together with  \eqref{Helm}, \eqref{Nbc}, \eqref{Helm S} and \eqref{Nbc S}, we find
\begin{equation}\label{rec scat}
(\kappa^2-k^2)\int_{\mathcal{D}_\lambda\!\!}v^*u\,dA - \int_0^\pi\lambda\left(v^*\pd{u}{r}-u\pd{v^*}{r}\right)_{r=\lambda}\,d\phi=0.
\end{equation}
Substituting the decomposition \eqref{decompose}, we can write \eqref{rec scat} in the form
\begin{equation}\label{rec scat modified}
(\kappa^2-k^2)\int_{\mathcal{D}_\lambda\!\!}v_s^*u\,dA - \int_0^\pi\lambda\left(v_s^*\pd{u}{r}-u\pd{v_s^*}{r}\right)_{r=\lambda}\,d\phi= \mathcal{F},
\end{equation}
where we define the `forcing' term
\begin{equation}\label{F def}
\mathcal{F}=\int_0^\pi\lambda\left(v_a^*\pd{u}{r}-u\pd{v_a^*}{r}\right)_{r=\lambda}\,d\phi-(\kappa^2-k^2)\int_{\mathcal{D}_\lambda\!\!}v_a^*u\,dA,
\end{equation}
which is linear in the ambient field. 

\subsection{Resonance}
In analysing the scattering problem, we shall restrict our attention to studying the resonant excitation of the anti-symmetric quasi-BIC in the limit $\epsilon\ll1$. By resonant excitation we mean that throughout the  domain the scattered field can be expanded as
\begin{equation}\label{res relation}
v_s= \mathcal{A} u+\cdots \quad \text{as} \quad \epsilon\to0,
\end{equation}
where $\mathcal{A}$ is a complex amplitude which remains to be determined and is expected to depend on the geometric parameters $\epsilon$ and $l$ as well as the mode number $n$. Given that $k''$ is exponentially small for the quasi-BIC, we expect such resonance excitation to occur in exponentially narrow wavenumber intervals. That is, for $\kappa$ such that  
\begin{equation}\label{res regime}
\kappa-k'=o\left(\epsilon^m\right) \quad \text{as} \quad \epsilon\to0
\end{equation}
for all real $m$. Our analysis will confirm the existence of these resonance intervals and uncover their precise exponential scaling.

We need to asymptotically estimate the integrals appearing in \eqref{rec scat modified} and \eqref{F def} in the limit $\epsilon\to0$, with $\kappa$ satisfying \eqref{res regime} and for some mode number $n$. As in \S\ref{ssec:calckpp}, we choose $1\ll\lambda\ll1/\epsilon$. Then, using \eqref{res relation}, leading-order approximations for the two integrals in \eqref{rec scat modified} follow from \eqref{num as} and \eqref{denom as}, respectively, up to the multiplicative factor $\mathcal{A}^*$. Together with \eqref{kpp sol} and \eqref{res regime}, these results give 
\refstepcounter{equation}
$$
\label{scat integrals}
\int_{\mathcal{D}_\lambda\!\!}v_s^*u\,dA\sim \epsilon  \mathcal{A}^*, \quad  \int_0^\pi\lambda\left(v_s^*\pd{u}{r}-u\pd{v_s^*}{r}\right)_{r=\lambda}\,d\phi\sim -4i k_0 k''\times \epsilon \mathcal{A}^* \quad \text{as} \quad \epsilon\to0.
\eqno{(\theequation{\mathrm{a},\!\mathrm{b}})}
$$
Moving to the integrals appearing in \eqref{F def}, we note that the area integral is order unity, hence negligible relative to (\ref{scat integrals}a) since for \eqref{res relation} to be valid in the exterior region  $\mathcal{A}$ must be asymptotically large. In that case, $\mathcal{F}$ is dominated by the first integral in \eqref{F def}, which can be approximated using \eqref{exterior exp}, the exterior expansion for $u$. Thus,  
\begin{equation}\label{Fint step 1}
\mathcal{F} \sim \epsilon^2\exp\left(-\frac{\pi l}{2\epsilon}\right)\times \int_0^\pi\lambda\left(v_a^*\pd{u_0}{r}-u_0\pd{v_a^*}{r}\right)_{r=\lambda}\,d\phi \quad \text{as} \quad \epsilon\to0. 
\end{equation}
To evaluate the latter integral, we use Jacobi-Anger expansions \cite{Abramowitz:book} to write the ambient field \eqref{ambient} as the Fourier series
\begin{equation}
v_a=2J_0(\kappa r)+4\sum_{m=1}^{\infty}\Lambda_mJ_m(\kappa r)\cos m \phi, 
\end{equation}
where $J_m$ denotes the Bessel function of order $m$ and
\begin{equation}\label{Lambda def}
\Lambda_m=\chi_m\cos m\alpha + i(1-\chi_m)\sin m\alpha,
\end{equation}
$\chi_m$ being unity for $m$ even and zero for $m$ odd. Since $u_0$ involves only the $\cos\phi$ Fourier mode, and given the orthogonality properties of the Fourier modes, only the term $4\Lambda_1J_1(\kappa r)\cos\phi$ in $v_a$ contributes to the integral on the right-hand side of \eqref{Fint step 1}, with \eqref{Lambda def} giving $\Lambda_1=i\sin\alpha$. Using \eqref{u0 far} and  \eqref{res regime}, in conjunction with the large-argument behaviour of $J_1$  \cite{Abramowitz:book}, we find
\begin{equation}\label{F asym}
\mathcal{F}  \sim\frac{16ik_0^2\sigma_n\sin\alpha}{\pi e}\epsilon^2 \exp\left(-\frac{\pi l}{2\epsilon}\right) \quad \text{as} \quad \epsilon\to0. 
\end{equation}

Substituting \eqref{scat integrals} and \eqref{F asym} into the reciprocal relation \eqref{rec scat modified}, we find the requisite asymptotic amplitude-wavenumber relation
\begin{equation}\label{A sol}
\mathcal{A}= -\frac{8i\epsilon\sigma_nk_0\sin\alpha}{\pi  e}\frac{\exp\left(-\frac{\pi l}{2\epsilon}\right)}{\kappa-k'-ik''}.
\end{equation}
In the denominator, it is not allowed to assume a leading approximation for $k'$, or even any finite number of terms in its expansion; this is because $\kappa$ and $k'$ cancel out up to the exponentially small $\epsilon^3\exp(-\pi l/\epsilon)$ order of $k''$. Since $v_a\simeq 1$ in the exterior region, and given \eqref{A sol} and the scaling of $k''$, the resonance condition \eqref{res relation} requires 
\begin{equation}\label{refined res regime}
\kappa-k' =\mathcal{O}\left(\epsilon^3e^{-\tfrac{\pi l}{\epsilon}}\right). %\exp\left(-\frac{\pi l}{\epsilon}\right)\right).
%O\left(\epsilon^3\exp\left(-\frac{\pi l}{\epsilon}\right)\right).
\end{equation}
This refines our initial estimate \eqref{res regime} for the resonance regime, in which \eqref{res relation} holds with $\mathcal{A}$ given by \eqref{A sol}. Note that in that regime the scaling of $\mathcal{A}$ is exponentially large: 
\begin{equation}
\mathcal{A}\simeq \frac{1}{\epsilon^2}\exp\left(\frac{\pi l}{2\epsilon}\right).
\end{equation}
In particular, at $\kappa=k'$, namely at resonance, $\mathcal{A}$ attains the value (cf.~\eqref{kpp sol})
\begin{equation}\label{A sol res}
\mathcal{A}\ub{r} = -\frac{\sigma_n \pi e \sin\alpha}{\epsilon^2k_0^2}\exp\left(\frac{\pi l}{2\epsilon}\right).
\end{equation}
\begin{figure}[t!]
\begin{center}
\includegraphics[trim=40 0 0 0, scale=0.36]{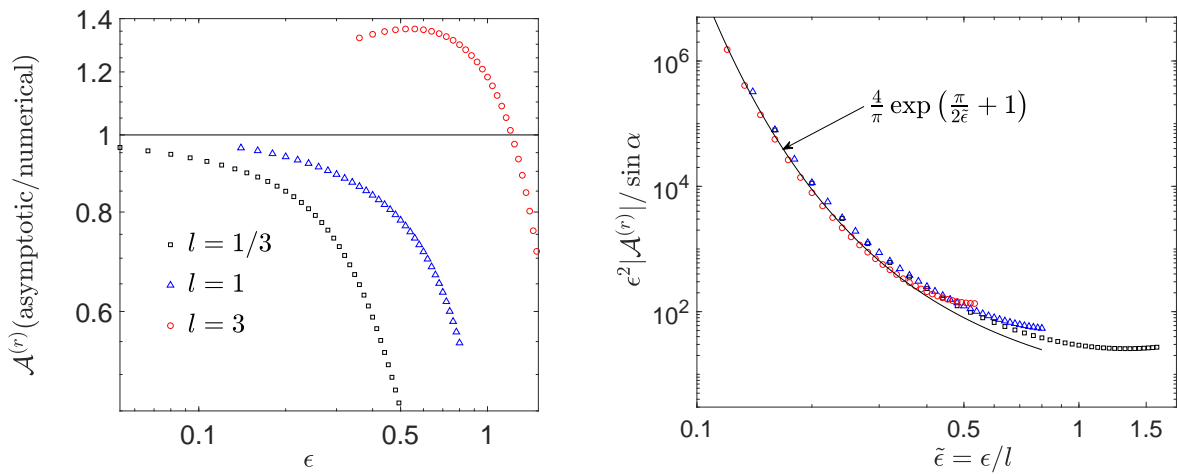}
\caption{The amplitude at resonance $\mathcal{A}\ub{r}$ for incidence angle $\alpha=\pi/4$. Left: ratio of the asymptotic approximation \eqref{A sol res} to numerically computed values as a function of $\epsilon$ and for the indicated values of $l$. Right: $\mathcal{A}\ub{r}$ and $\epsilon$ scaled so as to collapse numerical data for different $l$ on a single analytical curve implied by \eqref{A sol res}.}
\label{fig:Ar}
\end{center}
\end{figure}

The resonance condition \eqref{res relation}, together with the asymptotic amplitude-wavenumber relation \eqref{A sol} and the matched asymptotic expansions for the quasi-BIC eigenfunctions allows calculating a leading-order asymptotic approximation for the scattered field in all regions. In particular, from \eqref{part expansion} and \eqref{U0 sol} we see that $|\mathcal{A}|$ directly corresponds to the magnitude of the standing wave in the partitioned channel, which is accordingly exponentially large at resonance. Furthermore, using \eqref{exterior exp} and \eqref{u0 sol} we see that in the exterior region the scattered field possesses the asymptotic approximation
\begin{equation}\label{vs exterior result}
v_s \sim \frac{32k_0^3 \epsilon^3\exp\left(-\pi l/\epsilon\right)\sin\alpha}{\pi^2e^2(\kappa-k'-ik'')}{H}\ub{1}_1(k_0r)\cos\phi \quad \text{as} \quad \epsilon\to0.
\end{equation}
At resonance, substituting \eqref{kpp sol} shows that \eqref{vs exterior result} reduces to
\begin{equation}\label{vs res sol}
v_s\ub{r} \sim -4i\sin\alpha{H}\ub{1}_1(k_0r)\cos\phi \quad \text{as} \quad \epsilon\to0.
\end{equation}
Notably, the magnitude $4\sin\alpha$ of the scattered dipole at resonance is independent of both $l$ and $\epsilon$. The fact that the scattering magnitude at resonance approaches a constant as $\epsilon\to0$, namely as the scatterer size vanishes, means that the scatterer becomes increasingly efficient in that limit. %An energy argument similar to that in \cite{Schuller:09} and \cite{Porter:21bounds}, modified to the present semi-infinite geometry, shows that the limiting magnitude value $4\sin\alpha$ corresponds to a generic energy limit.  

We finish this section by comparing the above asymptotic approximations with numerical results. In our numerical scheme, the field in the right-hand side, say, of the partitioned portion of the slit is expanded in terms of eigenfunctions, the fundamental eigenfunction being the $x$-independent standing wave $\cos(\kappa(y+1+l))$; at a quasi-BIC resonance, this fundamental eigenfunction dominates the numerical solution. From \eqref{res relation} and \eqref{U0 sol}, the corresponding asymptotic approximation at resonance for the scattered (and total) field in the partitioned-channel region is $\mathcal{A}\ub{r}\cos(k_0(y+1+l))$. Given \eqref{res regime}, it is accordingly meaningful to compare $\mathcal{A}\ub{r}$ directly with the coefficient multiplying the fundamental eigenfunction in the numerical scheme. This comparison is shown in Fig.~\ref{fig:Ar}. Instead comparing $\mathcal{A}\ub{r}$ with the numerical wave field averaged over the right-hand side of the bottom boundary of the slit gives results which are indistinguishable from those presented.

\begin{figure}[t!]
\begin{center}
\includegraphics[scale=0.35]{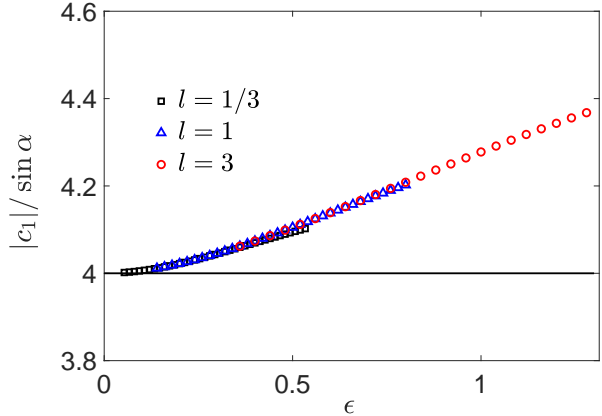}
\caption{The magnitude at resonance of the dipole coefficient $c_1$ in the multipole expansion \eqref{vs multipole} of the scattered wave for incidence angle $\alpha=\pi/4$, plotted as a function of $\epsilon$. Symbols: numerical values for indicated values of $l$. Solid line: the $l$-and $\epsilon$-independent asymptotic value $4$ (cf.~\eqref{vs res sol}).}
\label{fig:dipole}
\end{center}
\end{figure}

Furthermore, we can extract from our numerical scheme the coefficients $c_m$ in a conventional multipole expansion of the scattered far-field, 
\begin{equation}\label{vs multipole}
v_s=\sum_{m=0}^{\infty}c_m{H}\ub{1}_m(\kappa r)\cos m\phi,
\end{equation}
where ${H}\ub{1}_m$ denotes the $m$-th order Hankel function of the first kind and $c_m$ are diffraction coefficients. We refer to $c_0$ and $c_1$ as the monopole and dipole coefficients, respectively. Our asymptotic results imply that near the quasi-BIC resonances the dipolar term in \eqref{vs multipole} is dominant, with an asymptotic approximation for $c_1$ following from \eqref{vs exterior result}. As shown in Fig.~\ref{fig:dipole}, we have confirmed numerically that $|c_1|/\sin\alpha$ approaches $4$ at resonance as predicted by \eqref{vs res sol}. For small $\epsilon$, the remaining coefficients are, as expected, small when evaluated at the quasi-BIC resonances. 

%As further discussed in the concluding section, the remaining coefficients in the multipole expansion \eqref{} are negligible at resonance, except when symmetric and anti-symmetric resonances happen to coincide in which case the monopole coefficient $c_0$ may also be order unity. 

\section{Concluding remarks}\label{sec:conc}
\begin{figure}[t!]
\begin{center}
\includegraphics[scale=0.35]{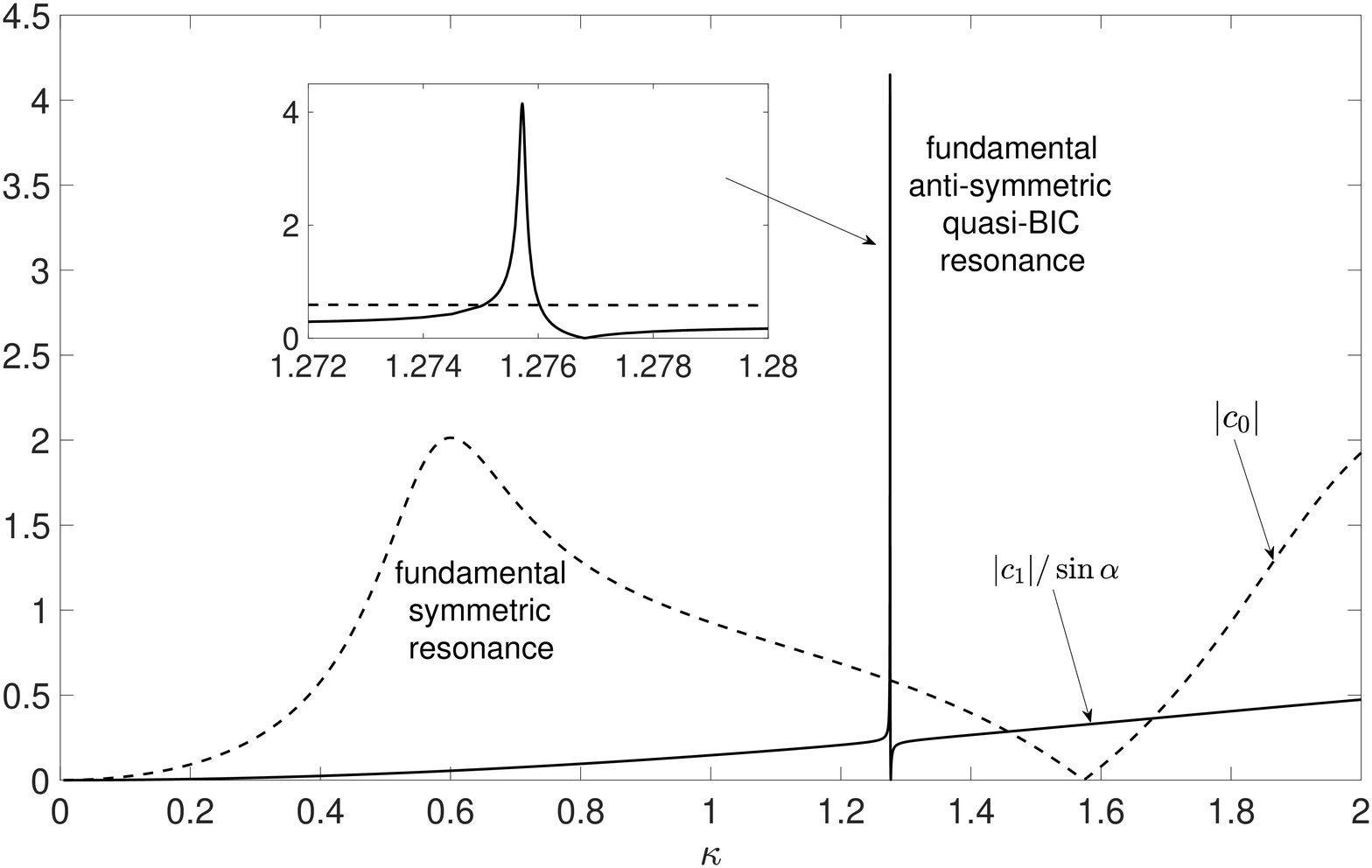}
\caption{Numerically calculated monopole and dipole coefficients in the multipole expansion \eqref{vs multipole} for the scattered field in the exterior region, for $\alpha=\pi/4$, $\epsilon=0.5$ and $l=1$. A moderate value of $\epsilon$ is chosen so as to be able to depict the exponentially narrow quasi-BIC resonance.} %Dashed curve: magnitude $|c_0|$ of the monopole coefficient. Solid curve: magnitude $|c_1|$ of the dipole coefficient, divided by $\sin\alpha$.}
\label{fig:response}
\end{center}
\end{figure}

Our analysis of the setup shown in Fig.~\ref{fig:dimsketch} has exclusively focused on the anti-symmetric quasi-BIC states and their resonant excitation. To put these resonances in a wider perspective, we plot in Fig.~\ref{fig:response} the first two diffraction coefficients in the multipole expansion \eqref{vs multipole}, calculated numerically, as a function of the wavenumber $\kappa$, for $\alpha=\pi/4$, $l=1$ and $\epsilon=0.5$. The quasi-BIC resonances correspond to the extremely narrow resonance peak in the anti-symmetric cylindrical-dipolar response. They are closely followed by anti-resonances, a signature of so-called Fano-type resonances associated with the interaction between an isolated quasi-bound mode and a continuous-spectrum of propagating modes \cite{Hein:10}; the anti-resonances occur outside the exponentially narrow resonance interval and are therefore not captured by our asymptotic analysis. The symmetric cylindrical-monopolar response is seen to exhibit less narrowband resonances; these correspond to the conventional standing-wave resonances of a slit channel open on one side, which are not perturbed at all by the presence of the thin partition. These resonances occur at wavenumbers $\kappa /a\approx (\pi/(1+l))(1/2+\bar{n})$, where $\bar{n}=0,1,2,\ldots$. Closely related resonances have been studied asymptotically \cite{Lin:17,Evans:18,Holley:19,Brandao:20}, showing, up to logarithmic modulations, order $\epsilon$ wavenumber corrections associated with end effects, order $\epsilon$ bandwidth (i.e., order $1/\epsilon$ quality factor) and order $1/\epsilon$ field enhancement inside the slit. 

The $n$th anti-symmetric, i.e., quasi-BIC, and $\bar{n}$th symmetric, i.e., conventional resonances can be made to approximately coincide by tuning the geometric parameter $l$ to $2(\bar{n}-n)/(1+2n)$. (A more precise condition could be derived by including $\epsilon$-dependent wavenumber corrections associated with end effects.) This may be of interest in order to achieve multi-polar radiation in the exterior region. For $l\simeq 1$, the field inside the slit under such simultaneous excitation would be dominated by the exponentially excited anti-symmetric state, notwithstanding the symmetric and anti-symmetric states radiating waves of comparable magnitude in the exterior region. Note that simultaneously exciting the \emph{fundamental} symmetric and anti-symmetric states ($n=\bar{n}=0$) requires $l$ near zero; our analysis however fails for $l=\mathcal{O}(\epsilon)$ in which case the partition-edge, unified-channel region and aperture regions merge, thereby eliminating the evanescent decay underpinning the exponential nature of the quasi-BIC. That regime could be investigated using a methodology similar to that used herein; we would still expect quasi-BIC derived from the BIC of the associated semi-infinite geometry, however with the quality factor diverging  algebraically rather than exponentially as $\epsilon\to0$. 

Alternatively, an exponential quasi-BIC exhibiting multi-polar radiation could be achieved by displacing the partition away from the centreline of the slit. As long as the partition's thickness can be neglected, there are still perfect BIC in the associated semi-infinite geometry \cite{Evans:93}. The evanescent tail of these states, however, would now have both symmetric and anti-symmetric components, which would `turn on' cylindrical monopole and dipole waves, respectively, at the slit aperture. It is possible that the mixture of radiation modes could be controlled through the lateral displacement of the partition. 

We conclude by proposing several additional extensions. Thus, it would be relatively straightforward to adapt the methodology in the paper to study quasi-BIC resonances in closely related geometries, including transmission scenarios involving partitioned slits that are open at both ends, analogous three-dimensional configurations, multiple interacting partitioned slits and scaling up to metasurfaces. Physically, nonlinear and dissipation effects are both likely to be present within the partitioned portion of the slit given the exponential enhancement of the field there. In particular, the exponentially large quality factors that we have calculated are radiative quality factors, namely with dissipation neglected. Maximal dissipation is known to occur, however, under critical-coupling conditions, where the radiative and dissipative quality factors are comparable, suggesting that exponential quasi-BIC are sensitive to exponentially small loss levels. Dissipation could be included from first principles following the analysis in \cite{Brandao:20} of thermal and viscous boundary-layer effects on the symmetric resonances of a slit channel in a transmission scenario.

\textbf{Acknowledgements.} RP is grateful for the financial support provided by EPSRC under grant number EP/V04740X/1.
\bibliography{refs.bib}
\end{document}